\documentclass[12pt]{amsart}
\usepackage{amsaddr}
\usepackage[
	a4paper, 
	top=1in, 
	bottom=1in, 
	left=0.8in, 
	right=0.8in, 
]{geometry}
\usepackage{amsmath}
\usepackage[numbers,sort&compress]{natbib}
\usepackage{scrextend}
\usepackage{hyperref}
\usepackage{url}
\usepackage{booktabs}
\usepackage{multirow}
\usepackage{graphicx}
\usepackage{tabularx}
\usepackage{enumerate}
\usepackage{color}
\usepackage{amsmath}
\usepackage{bbding}
\usepackage{xcolor}
\usepackage{ifthen}
\makeatletter
\g@addto@macro{\endabstract}{\@setabstract}
\newcommand{\authorfootnotes}{\renewcommand\thefootnote{\@fnsymbol\c@footnote}}%
\makeatother
\usepackage{lineno}
\newcounter{is_nc}

\begin{document}

\begin{center}
 \LARGE 
 Leveraging Reaction-aware
Substructures for \\ Retrosynthesis Analysis \par \bigskip

  \normalsize
  \authorfootnotes
  Lei Fang\textsuperscript{1}\footnote{Corresponding author.}, 
  Junren Li\textsuperscript{2}\footref{footnote1},
  Ming Zhao\textsuperscript{3}\footnote{\label{footnote1}This work was done when Junren Li and Ming Zhao were interns at Microsoft Research Asia.},
  Li Tan\textsuperscript{4} and
  Jian-Guang Lou\textsuperscript{1} \par \bigskip

  \textsuperscript{1}Microsoft Research, \{leifa, jlou\}@microsoft.com \par
  \textsuperscript{2}Peking University, lijunren@pku.edu.cn \par 
  \textsuperscript{3}Waseda University, oldbulb@fuji.waseda.jp \par
  \textsuperscript{4}Mincui Therapeutix, li.tan@protonmail.com \par
  
  \bigskip

\end{center}

\setcounter{footnote}{0}

\begin{abstract}
Retrosynthesis analysis is a critical task in organic chemistry central to many important industries. 
Previously, various machine learning approaches have achieved promising results on this task by representing output molecules as strings and autoregressively decoded token-by-token with generative models.
Text generation or machine translation models in natural language processing were frequently utilized approaches. 
The token-by-token decoding approach is not intuitive from a chemistry  perspective because some substructures are relatively stable and remain unchanged during reactions.
In this paper, we propose a substructure-level decoding model, where the substructures are reaction-aware and can be automatically extracted with a fully data-driven approach.
Our approach achieved improvement over previously reported models, and we find that the performance can be further boosted if the accuracy of substructure extraction is improved.
The substructures extracted by our approach can provide users with better insights for decision-making compared to existing methods.
We hope this work will generate interest in this fast growing and highly interdisciplinary area on retrosynthesis prediction and other related topics.
\end{abstract}

\section{Introduction}\label{sec:intro}
Organic synthesis is an essential branch of synthetic chemistry that mainly involves the construction of organic molecules through various organic reactions. 
Retrosynthesis analysis\citep{corey1988robert} that aims to propose possible reaction precursors given a desirable product is a crucial task in computer-aided organic synthesis. 
Accurate predictions of reactants could help find optimized reaction pathways from numerous possible reactions.
Recently, machine learning-based approaches have achieved promising results on this task.
Many of these methods employ encoder-decoder frameworks, where the encoder part encodes the molecular sequence or graph as high dimensional vectors\citep{schwaller2019molecular,tetko2020state,duan2020retrosynthesis,wang2021retroprime,Seo_Song_Yang_Bae_Lee_Shin_Hwang_Yang_2021,irwin2022chemformer,tu2021permutation}, and the decoder attends to the output from the encoder and predicts the output sequence token by token in an autoregressive manner.
Note that the sequences of the molecules involved in these algorithms are usually represented as SMILES (Simplified Molecular-Input Line-Entry System) strings\citep{weininger1988smiles,weininger1989smiles}, and the graph refers to the molecular graph structure.
For example, \citet{tetko2020state} uses textual SMILES representations of reactants and products and formulates retrosynthesis analysis as a machine translation task from one language (product) to another (reactants) using data-augmented Molecular Transformer\citep{schwaller2019molecular}.

Casting retrosynthesis analysis as a machine translation task enables the use of deep neural architectures that are well developed in natural language processing.
For example, the self-attention based Transformer architectures\citep{vaswani2017attention} are employed in recent state-of-the-art  models\citep{schwaller2019molecular,tetko2020state,duan2020retrosynthesis,wang2021retroprime,irwin2022chemformer,Seo_Song_Yang_Bae_Lee_Shin_Hwang_Yang_2021}.
In the decoding stage, the output SMILES is autoregressively generated token-by-token.
It should be noted that this is not considered intuitive or explainable for chemists in synthesis design or retrosynthesis analysis. 
In real world route-scouting tasks, synthetic chemists generally rely on their professional experience to formulate a reaction pathway drawing inspirations from previously learned reaction pathways. 
Retrosynthesis analysis often starts from molecular substructures or fragments that are chemically similar to or inclusive in the target molecules.  
These substructures or fragments help provide clues to an assembly puzzle involving a series of chemical reactions toward the final product. 

\begin{figure}
    \begin{center}
        \includegraphics[scale=0.23]{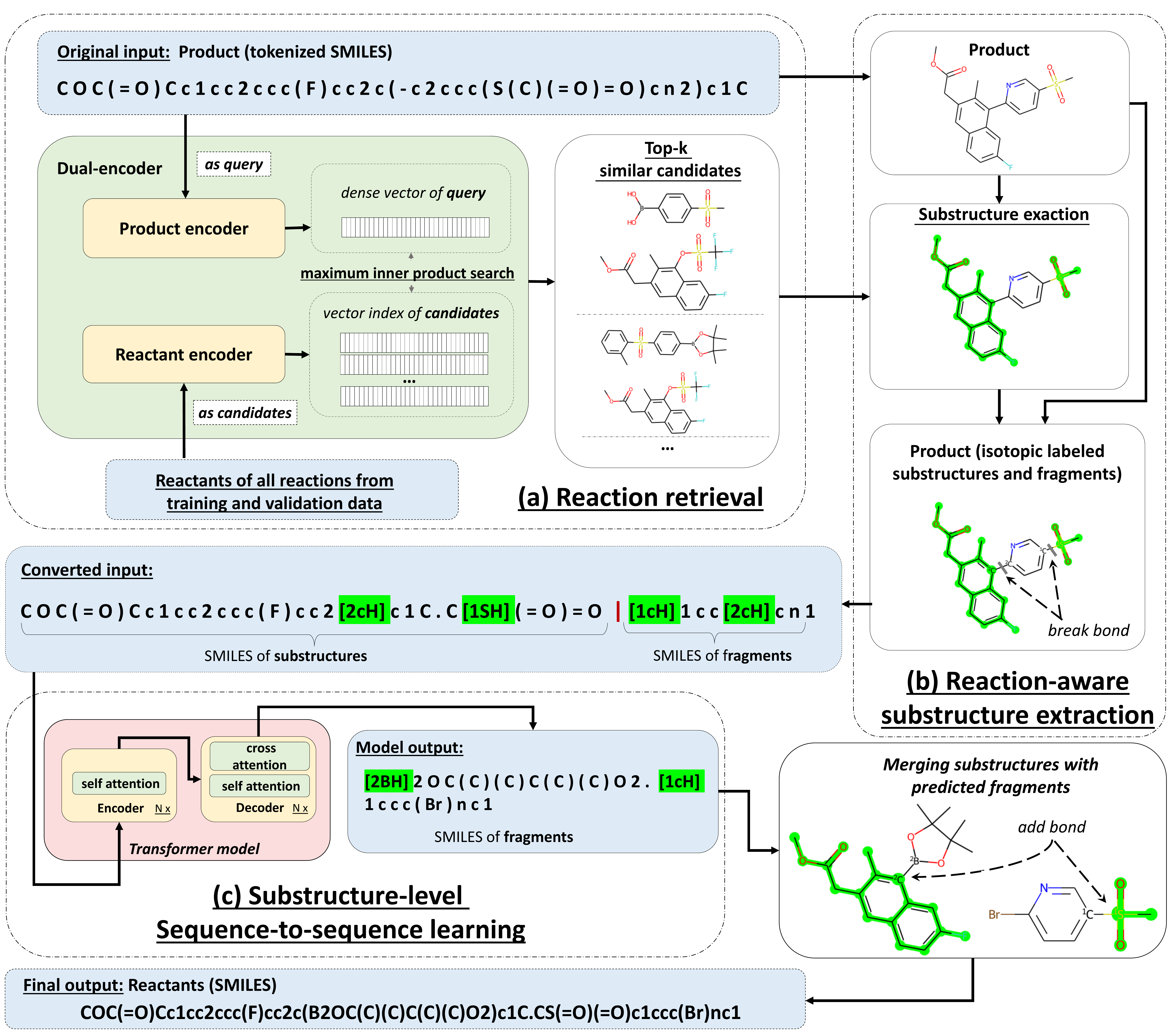}
    \end{center}
    \caption{Method overview.}
    \label{fig:model}
\end{figure}

In this paper, we propose to leverage reaction-aware substructures in organic synthesis, where
the substructures can help capture the subtle chemical changes among reactants and products, while remaining free from expert systems or template libraries.
We cast the retrosynthesis analysis as a sequence-to-sequence learning task with reaction-aware substructures.
The pipeline of the overall framework is illustrated in Figure~\ref{fig:model}, which consists of the following modules:
\begin{enumerate}[(a)]
    \item \textbf{Reaction Retrieval} \\
    The reaction retrieval module aims to retrieve similar reactions given the product as a query, and these reactions will be used for reaction-aware substructure extraction.
    We introduce a learnable cross-lingual memory retriever\citep{cai-etal-2021-neural} used in machine translation tasks to align the reactants and the corresponding products in high dimensional space. 
    The retrieval model is based on the dual-encoder framework\citep{bromley1993signature} such that for each reaction, the learned representation of reactants is similar to that of the product. 
    After the dual-encoder retrieval model is trained, we could obtain the dense vector representations of all the reactants and products, as is shown in Figure~\ref{fig:model}(a).
    In retrosynthesis analysis, the product will be the query to retrieve reactant molecules that are similar in the high dimensional space.
    For a fair comparison to other methods, the retrieved candidates only come from the training and validation data.
    \item \textbf{Reaction-aware Substructure Extraction} \\
    Given the training objective of the dual-encoder retrieval model, the retrieved molecules should be similar to the golden target (the reactants).
    Therefore, we could extract the common substructures from the query molecules and the top cross-aligned candidates based on molecular fingerprints, and assume these common structures exist in the golden targets.
    Note that these substructures are relatively stable and remain unchanged during the reaction.
    More details are provided in Section~\ref{sec:substructure}. 
    The common substructure provides a reaction-level, fragment-to-fragment mapping from reactants to products.
    It should also be noted that these substructures are reaction-aware and could be considered as reaction templates learned from the dual retrieval model. 
    We then separate the molecules into common substructures and other molecular fragments.
    Molecular fragments in this paper refer to atoms and bonds that are not in the common substructure.
    When multiple bonds are broken to isolate the substructures, we introduce ``isotopic numbers'' to virtually tag the atoms of the broken bonds, as is shown in Figure~\ref{fig:model}(b).
    This method is analogous to using an isotope to label the atom of interest during a Nuclear Magnetic Resonance (NMR) study. 
    It is important to note that our isotopic number labels do not actually denote changes/differences in the number of neutrons in an atom’s nuclei.
    The isotopic number is not related to an atom’s chemical identity but is solely used to track sites of bond breaking.
    
    \item \textbf{Substructure-level Sequence-to-sequence Learning} \\
    With the reaction-aware substructure and molecular fragments, we convert the original token-level sequence to a substructure-level sequence.
    The new input sequence will be the SMILES of the substructure followed by the SMILES of other fragments with isotopic numbers.
    The output sequence will be the isotopically labeled fragments. In other words, the fragments are connected to common structures with bonds specified by isotopic labels. 
    Subsequently, retrosynthesis analysis will be cast to the structure-level sequence-to-sequence learning task.
    Given the model predicted isotopically labeled fragments, we perform bottom-up modular assembly of the molecular architectures to obtain the final molecular graph and its SMILES strings.
    An example is shown in Figure~\ref{fig:model}(c): with the model output sequence, \texttt{1S} (denoted by \texttt{[1SH]}) is an  isotopically labeled atom from the substructure, and it shall be attached to the atom \texttt{1c} (denoted by \texttt{[1cH]}) in the predicted fragment because they are labeled by the same isotopic number of \texttt{1}, and similarly, \texttt{2c} (denoted by \texttt{[2cH]}) from the substructure shall be connected to the atom \texttt{[2B]}(denoted by \texttt{[2BH]}) in the predicted fragment. 
    
\end{enumerate}
The substructures are intrinsically related to how human researchers interpret the nature of chemical reactions, and our approach achieved improvement over previously reported models.
We show that the performance can be further boosted if the accuracy of substructure extraction is improved. 
The substructures extracted by our approach can provide users with better insights for decision-making compared to existing methods.

\section{Methods}\label{sec:appoach}
\subsection{Reaction Retrieval}\label{sec:retri}
In reaction pathway planning, chemists generally need to
obtain insights and inspirations from existing reaction pathways learned through previous education and professional experience. 
Similarly, the retrieval module shall obtain a list of candidates similar to the given query from a large collection of data efficiently.
For retrosynthesis analysis, the query will be the product, and the candidates will be a list of reactants from ``existing'' reactions (training and validation data).
\begin{figure}
\begin{center}
    \includegraphics[scale=0.23]{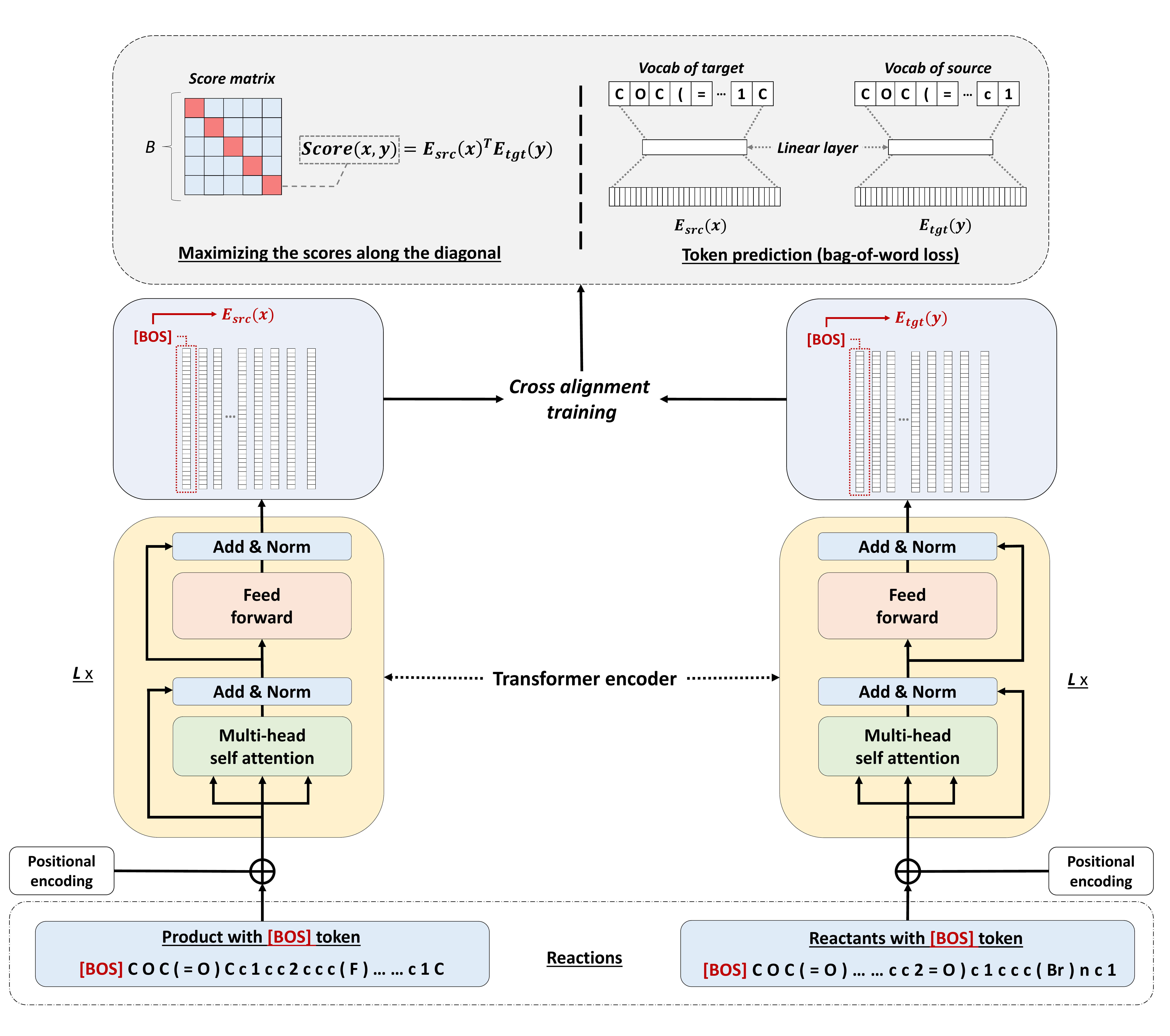}
\end{center}
    \caption{Overview of the dual-encoder retrieval model.}
    \label{fig:dual}
\end{figure}
To learn and measure the similarity between the reactants and the product, we use the dual-encoder architecture\citep{bromley1993signature}, which was also introduced prior in memory-based machine translation\citep{cai-etal-2021-neural}. 
We use two independent Transformer encoders\citep{vaswani2017attention} in the dual-encoder architecture. One is to encode the reactants, and the other is to encode the products, as shown in Figure~\ref{fig:dual}. 

Transformer\citep{vaswani2017attention} is a prominent encoder-decoder model that has achieved great success in natural language processing, computer vision, and speech processing. 
It consists of an encoder and a decoder, each of which is a stack of 
$L$ identical blocks, 
where each encoder block is mainly a combination of a self-attention module and a position-wise feed-forward network. 
Please refer~\citet{vaswani2017attention} for details about the Transformer model.
Note that we only employ the Transformer encoder in dual-encoder.

We add the \texttt{[BOS]} token to the tokenized SMILES strings of both products and reactants, which are then fed into the Transformer source and target encoders. 
The source and target representations of \texttt{[BOS]} token are considered as the output, denoted by $E_{\mathrm{src}}$ and $E_{\mathrm{tgt}}$, respectively.
The overall objective is to minimize the distance between $E_{\mathrm{src}}$ and $E_{\mathrm{tgt}}$ in high-dimensional space for a given reaction.
Following the training strategy proposed by~\citet{cai-etal-2021-neural}, we propose two objectives for cross-alignment. 
The first objective is that the gold target has the highest-ranking score given the source among all the targets.
This is approximated by maximizing the ranking score in a batch of source-target pairs when the batch size is relatively large.
For a batch of $B$ source-target pairs sampled from the training set at each training step, let $X$ and $Y$ be the $B\times d$ matrix of the encoded source and target vectors, respectively.
We define the ranking scores as the dot product of the encoded source and target representations. We have $S=X Y^{T}$, which is a $B \times B$ matrix of scores, where each row corresponds to one source, and each column corresponds to one target in the batch.
The pair $\left(X_{i}, Y_{j}\right)$ should be aligned when $i = j$, and otherwise not. 
The goal is to maximize the scores along the diagonal of the matrix and henceforth reduce the values in other entries.
The loss function for the $i$-th source-target pair is as follows:
\begin{equation}
    \mathcal{L}_{\mathrm{rank}}^{(i)}=\frac{-\exp \left(S_{i i}\right)}{\exp \left(S_{i i}\right)+\sum_{j \neq i} \exp \left(S_{i j}\right)}
\end{equation}
The second objective is mainly borrowed from machine translation, which aims to predict the tokens in the target given the source representation and vice versa. 
This objective introduces additional semantic alignment between source and target at the token level. 
For the $i$-th source-target pair, the bag-of-words loss is used for this token-level cross-alignment and is formulated as:
\begin{equation}
    \mathcal{L}_{\text {token }}^{(i)}=-\sum_{w_{y} \in \mathcal{Y}_{i}} \log p\left(w_{y} \mid X_{i}\right)+\sum_{w_{x} \in \mathcal{X}_{i}} \log p\left(w_{x} \mid Y_{i}\right)
\end{equation}
where $\mathcal{X}_{i}$ and $\mathcal{Y}_{i}$ denote the set of tokens in the $i$-th source and target, respectively.
The probability $p$ is computed by a linear projection layer followed by a softmax layer. 
For the dual-encoder model, the overall loss is:
\begin{equation}
\mathcal{L} = \frac{1}{B} \sum_{i=1}^{B} \mathcal{L}_{\mathrm{rank}}^{(i)}+\mathcal{L}_{\mathrm{token}}^{(i)}.
\end{equation}

When the dual-encoder is trained, we can obtain the dense vectors for all the targets (the reactant side of reactions) in the training and validation data.
We leverage Faiss\citep{johnson2019billion}, an open-source toolkit\footnote{\url{https://github.com/facebookresearch/faiss}}, to perform Maximum Inner Product Search (MIPS) on large collections of dense vectors. 
It does so by building the index of dense vectors, which is optimized for MIPS search.
The Faiss index code in our work is ``IVF1024 HNSW32, SQ8'', which is the graph-based index with Hierarchical Navigable Small World (HNSW) algorithm\citep{malkov2020efficient}.
In our approach, we pre-compute and index the dense vector representations of all targets on the training and validation data with the target encoder. 
For input query $x$, which is the product in retrosynthesis analysis, we use the
source encoder to obtain its dense vector representation  $E_{\mathrm{src}}(x)$, and retrieve a ranked list of candidates by MIPS on the Faiss index.

\subsection{Reaction-aware Substructure Exaction}
\label{sec:substructure}
Given the training objective of the dual-encoder model, the retrieved top candidates shall be similar to the golden target.
We further assume that these candidates share a common substructure with the golden target.
Although this hypothesis is not always valid, we observe that the assumption is reasonable in most cases.
These common substructures are reaction-aware because the retrieved candidates vary for different reactions.
Our goal is to extract reaction-aware substructures given the query and the top cross-aligned targets obtained.

The extraction is mainly based on molecular fingerprint, which is widely used in molecular substructure and similarity search.
Molecular fingerprints are a way of encoding the structure of a molecule. The most common type of fingerprint is a series of binary bits that represent the presence or absence of particular substructures in the molecule. 
Comparing fingerprints can help determine the similarity between two molecules or locate the aligned atoms.
Circular fingerprints are one of the methods capable of capturing 3D topological information. It maintains the environment of the center atom, which covers the neighbor atoms in different radii.
The \textit{de facto} standard circular fingerprints are the Extended-Connectivity Fingerprints (ECFPs), based on the Morgan algorithm, which is specifically designed for structure-activity modeling. 
Circular fingerprints are obtained through an enumeration of sub-molecular neighborhoods. 
First, each atom is encoded by an integer identifier, which is a hashed encoding representation of structural properties. 
The neighborhood information of the constituent atoms and bonds in different radii are iteratively assigned as the atom's numerical identifiers. 
The radius of a circular fingerprint refers to the size of the largest neighborhood surrounding each atom that is considered during enumeration.
The fingerprint consists of the combination of all unique identifiers and is subsequently folded into a binary vector of fixed length by converting integer identifiers into indices of the vector. 
\begin{figure}[ht]
    \centering
    \includegraphics[scale=0.23]{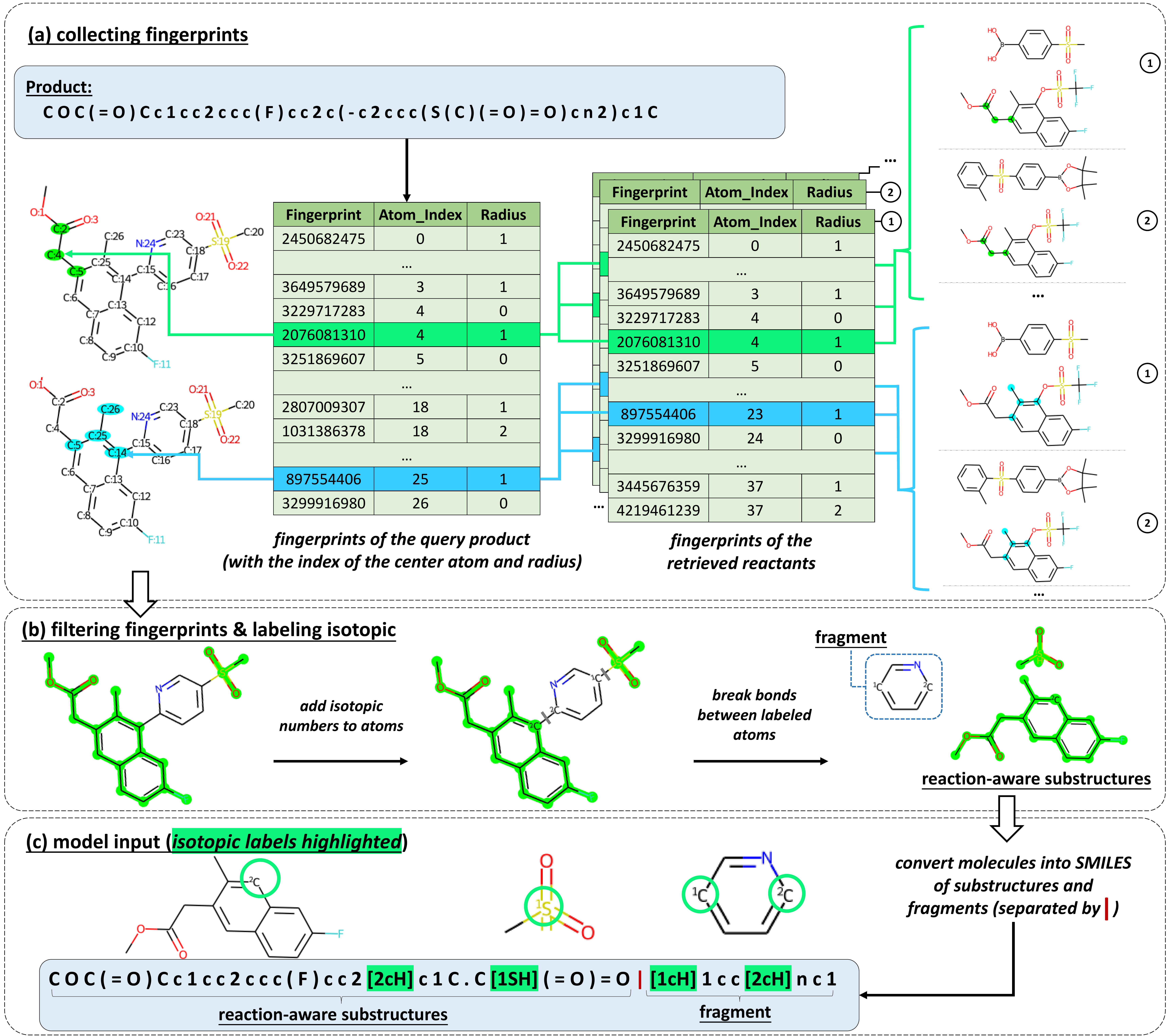}
    \caption{Reaction-aware substructure extraction.}
    \label{fig:sub}
\end{figure}
We use the toolkit RDKit\citep{greg_landrum_2022_6388425} to extract common substructures.
The overall extraction scheme is illustrated in Figure~\ref{fig:sub}. 
In our approach, we calculate the circle fingerprints of the query and the top $20$ retrieved candidates with a radius ranging from $2$ to $6$.
For example, in Figure~\ref{fig:sub}(a), the fingerprint \texttt{2076081310} encodes the environment of the center atom (index \texttt{4}) and its neighbors in radius \texttt{1} in the query.

For each candidate, we build the atom alignments with the query using the shared fingerprints, as are highlighted in green and blue in the fingerprint table in  Figure~\ref{fig:sub}(a).
We select atoms to build the substructure if they are aligned $5$ times or more among the retrieved candidates. 
We further remove atoms in the substructure that are aromatically bond to non-substructure atoms, or the connecting bonds show stereoisomerism. 
Otherwise, separating molecules into fragments might be counterintuitive from a chemistry perspective, e.g., it may destroy the aromaticity or stereoisomerism of the original molecule. 
We will explain how we handle these removed atoms that remain unchanged during reaction in Section~\ref{sec:transformer}.
For simplicity, we also remove atoms that are connected to multiple non-substructure atoms. Note that for a specific reaction, the atoms in the extracted substructure might not be fully connected. 
They could be different parts of one molecule or parts of different molecules, as is shown in Figure~\ref{fig:sub}(b).

Now we separate the query into substructures and other fragments.
The assumption is that these substructures remain unchanged during the reaction.
Note that we might have multiple fragments connected to atoms of the substructure. 
We introduce isotopic numbers as labels to differentiate these bonds.
As is shown in Figure~\ref{fig:sub}(b), we add the isotopic label to the bond between the atom \texttt{S} in the common substructure, and the atom \texttt{c} in the fragment, resulting in SMILES snippets with the isotopic labels \texttt{[1SH]} and \texttt{[1cH]}, respectively.
Note that we introduce additional hydrogen atoms in the substructure and other fragments after breaking the bonds, making them look like charge-neutral molecules rather than radicals. 
We also record the type of breaking bond type (double or triple) so that we can remove these hydrogen atoms easily when restoring the original molecule.
Atoms with same isotopic number means that they are connected in the original molecule, for example, \texttt{[1SH]} is connected to \texttt{[1cH]} and \texttt{[2cH]} in the substructure is connected to the atom \texttt{[2cH]} in the fragment.
With isotopic number labeling the bonds, we can easily isolate the substructure from other molecule fragments or restore the molecule from the substructure and other fragments.
The broken bonds between the substructure and other fragments do not necessarily mean that they will be the sites of reactivity. 
Only some of the broken bonds might become reactivity sites, for example, the breaking bond between \texttt{1S} and \texttt{1c} is not the site of reactivity in Figure~\ref{fig:sub}.

To build a model which is not sensitive to the substructure for a given product molecule, we also extract the center and neighbor atoms based on the common fingerprints as substructures from all the retrieved candidates.
These substructures may be different as they are from different candidates.
All the substructures from retrieved candidates that exist in the query (product) will be used as input for model training and inference.
In other words, the product molecule will be represented multiple times with different substructures and fragments.
In this way, the model is expected to output robust results over different substructures for a specific input product molecule.
By doing this, we could also group the predictions by substructures, which provide the user with more insights for decision-making in retrosynthesis planning compared with existing ``black-box'' models.

\subsection{Substructure-level sequence-to-sequence learning}\label{sec:transformer}
We can also isolate the substructure on the target side of the training data.
The source and target molecules are both converted into substructures and other molecular fragments. 
We use the SMILES strings to represent these substructures and fragments and cast retrosynthesis analysis as substructure-level sequence-to-sequence learning problems.
For sequence-to-sequence learning-based approaches, Molecular Transformer\citep{schwaller2019molecular,tetko2020state} has achieved state-of-the-art performance on the reaction outcome prediction and retrosynthesis analysis\citep{duan2020retrosynthesis}, it uses textual SMILES representations
of reactants, reagents, and products, and treat reaction prediction or retrosynthesis as a machine translation task.
The output SMILES is decoded by a Transformer decoder token-by-token.

\begin{figure}[ht]
    \centering
    \includegraphics[scale=0.18]{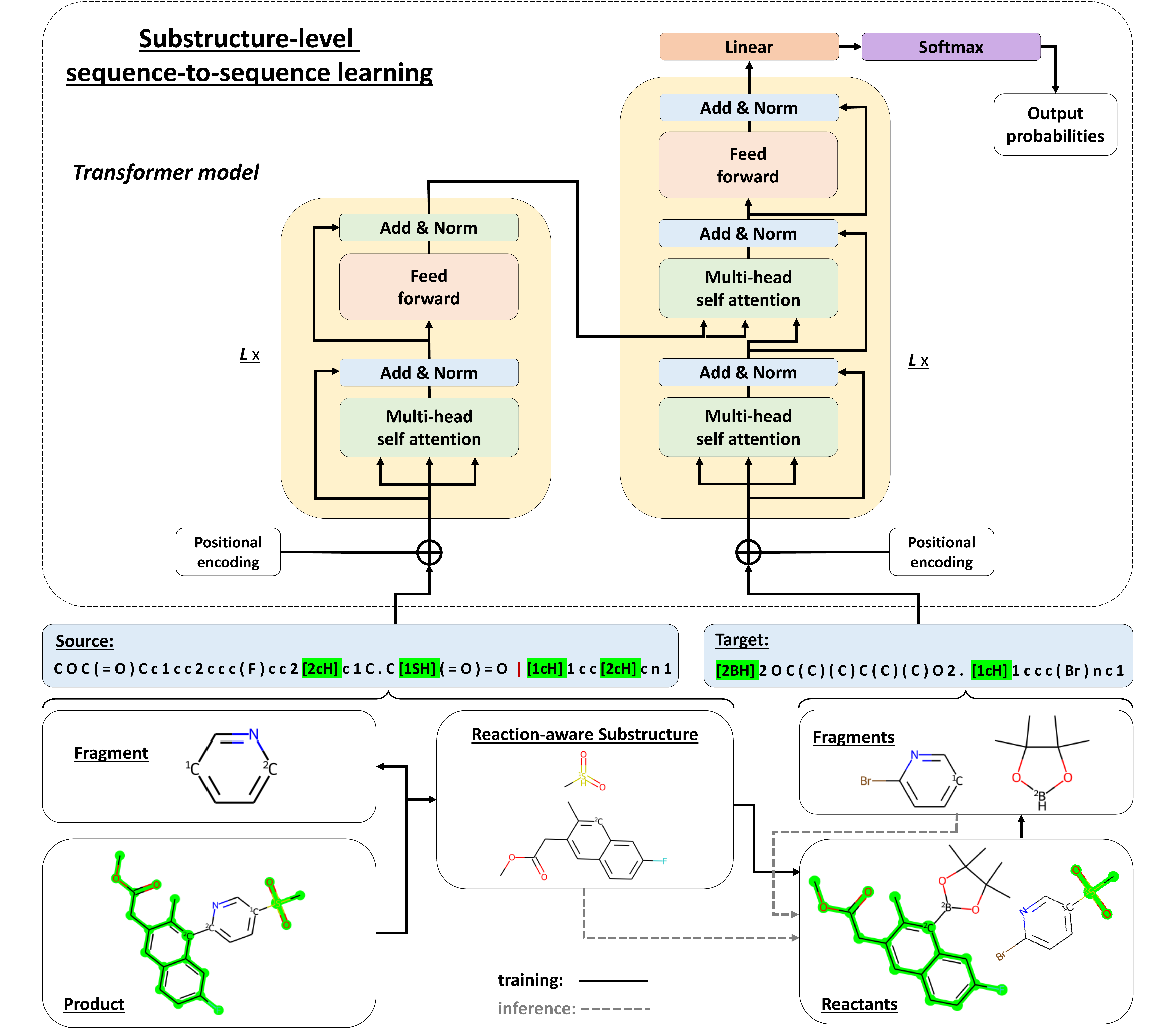}
    \caption{Substructure-level sequence-to-sequence learning.}
    \label{fig:s2s}
\end{figure}

In our approach, the input sequence will be the SMILES of the substructure and fragments separated by ``$|$'', as shown in Figure~\ref{fig:model} and Figure~\ref{fig:s2s}.
We assume that the substructure is stable and remains unchanged during the reaction.
For the target, we only need to predict the isotopically labeled fragments.
In case we failed to extract any substructures for a given query and the retrieved targets because there might be no atoms with the number of fingerprint alignments above the threshold (the minimal number of alignments is $5$ out of the $20$ retrieved targets), the input will fall back to SMILES, and will be predicted with a data augmented Transformer model\cite{tetko2020state}.
Based on our formulation, the task of retrosynthesis analysis is simplified, and the average length of the target sequences is significantly reduced compared to earlier models, which also helps to reduce the model complexity.
With extracted substructures and predicted fragments containing isotopic numbers, we could easily obtain the target molecules, which will be the predicted reactants, as shown in Figure~\ref{fig:model}.
The chemical changes among reactants and products are expected to be captured and predicted by the substructure-level sequence-to-sequence learning model.
Note that for those atoms that remain unchanged during reactions but are not included in the substructures, it will be predicted by the model in the output fragments.

Given an input product molecule, we extract substructures from all the retrieved candidates.
The original product molecule will be represented differently multiple times, each time with one substructure and the corresponding fragments.
During inference, different substructures may lead to the same reactant molecules with different rankings.
\citet{tetko2020state} calculated the ranking score for the predictions of augmented SMILES mainly based on the rank output by the beam search.
However, in our approach, the lengths of predicted fragments for different substructures are different.
It is not easy to define an empirical ranking formula based on the beam search rank. 
Therefore, we train a pair-wise ranking model, which is a neural network with three linear layers, on the validation data with input features such as the frequency, the ratio of ranking among top $1$ and top $2$, average rankings on predictions of all the substructures, and unique substructures.
The training objective is to ensure that on the validation data, the golden target has a higher score than incorrect predictions.

\section{Results}\label{sec:experiment}
\subsection{Data \& Settings}
\label{sec:exp:data}
We use the publicly available reaction datasets USPTO\citep{lowe2012extraction}, which use the SMILES strings to describe the chemical reactions.
We test our approach on the USPTO\_full benchmark with the same data split (train/valid/test set to 80\%/10\%/10\%) following~\citet{dai2019retrosynthesis}\footnote{We do not evaluate our approach on the USPTO\_50k~\cite{dai2019retrosynthesis}, which is a subset of the USPTO\_full, because the size is limited to obtain reasonably good substructures, and USPTO\_full is more challenging. }.
There are approximately 1M reactions in total.
We perform evaluations using the top-$k$ exact match accuracy, i.e., given a source, whether one of the $k$ generated targets exactly matches the ground truth. 
We canonicalize the molecules with the toolkit RDKit\citep{greg_landrum_2022_6388425} and tokenize all the inputs following~\citet{schwaller2018found}.
For each instance, we add two randomized SMILES as augmented data for substructures and fragments on the product side for both training and testing.

\begin{table}
\begin{center}
\begin{tabular}{@{}lcc@{}}
\toprule
Parameters & Dual-encoder & Substructure-level seq-to-seq        \\ \midrule
Embedding size             & 512           & 512                 \\
Hidden size                & 256           & 512                 \\
Feedforward hidden size   & 2048          & 2048                \\
Encoder blocks             & 3             & 10                  \\
Encoder attention heads    & 6             & 8                   \\
Total training steps       & 500,000       & 500,000             \\
Warm-up steps              & 4,000         & 8,000               \\
Learning rate              & 0.0001        & 0.001               \\
Dropout                    & 0.1           & 0.1                 \\ \bottomrule
\end{tabular}
\end{center}
\caption{Transformer parameter settings in the dual-encoder and the substructure-level sequence-to-sequence model. }
\label{tab:para}

\end{table}

Table~\ref{tab:para} shows the parameter settings.
For the dual-encoder reaction retrieval model, we follow the same parameter settings described in~\citet{cai-etal-2021-neural}. 
The batch size is $4096$, and the label smoothing is $0.1$. 
For other parameters in substructure-level sequence-to-sequence learning, we mainly follow the Molecular Transformer settings~\citep{schwaller2019molecular}.
We use the adam optimizer\citep{DBLP:journals/corr/KingmaB14} (${\beta}_1 = 0.9$, ${\beta}_2 = 0.998$) and the same Noam learning rate scheduler as described in\citep{vaswani2017attention}. 
The ranking model has three linear layers of size 400 and is trained on the pairs obtained on the validation data using the label smoothed cross-entropy loss. 
For fair comparisons, we also train a Transformer model with data augmentation ($5$ random + $1$ canonicalized SMILES at the product side) under the same settings to obtain predictions for product molecules with no substructures extracted.

\subsection{Reaction-Aware Substructures}
\begin{figure}[ht]
    \begin{center}
    \includegraphics[scale=0.22]{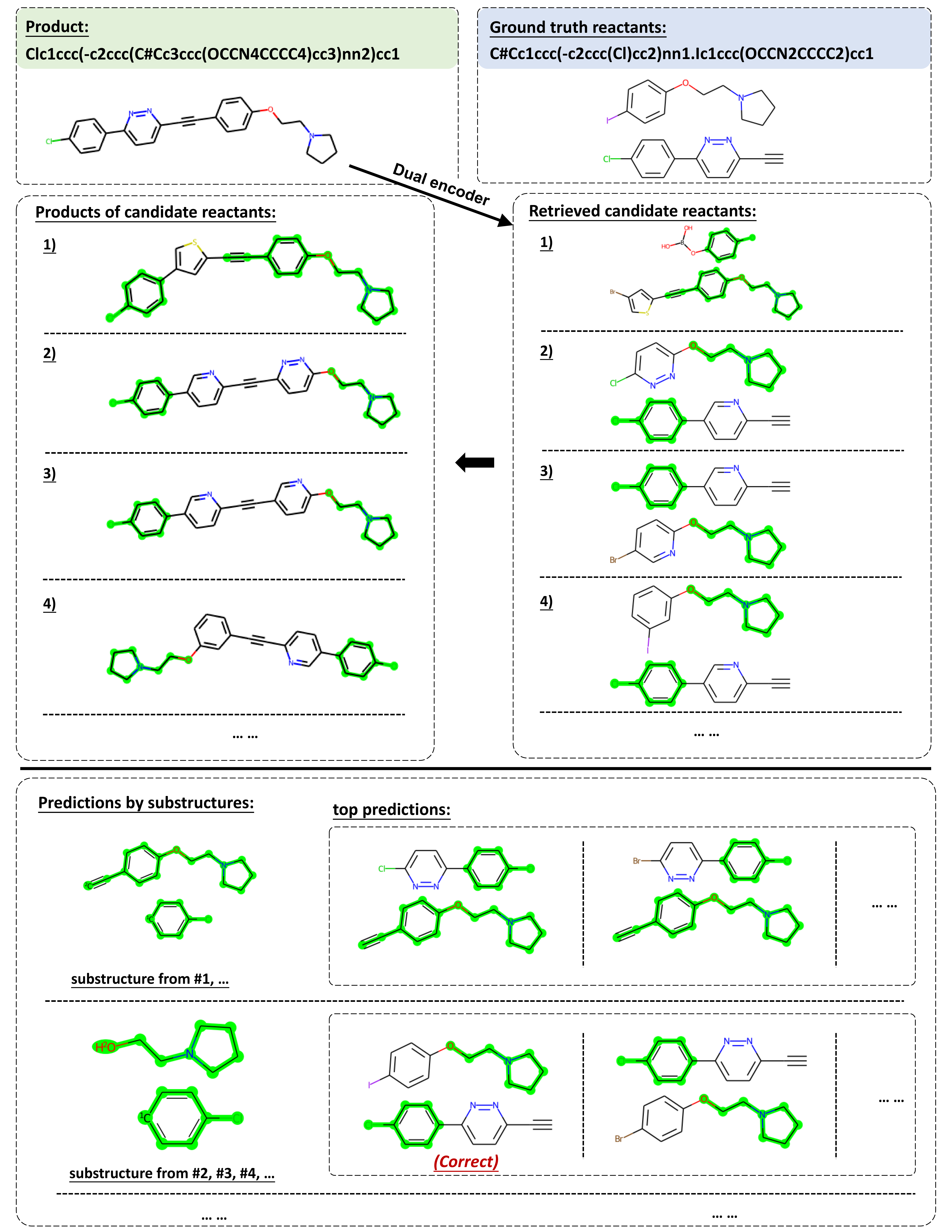}
    \caption{Predictions by substructures.}
    \label{fig:insight}
    \end{center}
\end{figure}
Note that the substructures are obtained based on the common fingerprints between the query product and the retrieved candidates,
and the retrieved candidates are not the golden reactants, this might introduce errors during substructure extraction.
The top half of Figure~\ref{fig:insight} shows an incorrect substructure extracted from candidate \#1.
The retrosynthesis product is a long molecule linked by a triple bond. 
The results show that all the retrieved candidates share common substructures with the product.
By taking a further look at the products of all the corresponding candidates, we readily observe that the triple bond could be the site of reactivity, which shall not have been included in the substructure, even if it is in the environment of the aligned fingerprint.
We leave this as future work to improve the accuracy of extracted substructures, i.e., we planned to identify possible sites of reactivity based on the retrieved candidates and exclude those atoms from substructures.

For incorrect substructures, we can easily filter them out with golden reactants on the training and validation data. 
On the training data, we extract substructures for $81.9\%$ product molecules after filtering incorrect substructures, the average number of substructures is $12.5$ (note that we extract substructures from all the retrieved candidates), and the average number of unique substructures is $4.2$.
We construct the model training data with unique substructures only.

On the test data, we extract substructures for $82.2\%$ products with an accuracy of $90.2\%$, and the average number of substructures and unique substructures is $12.1$ and $4.9$, respectively.
The average number of heavy atoms in the product, substructures, and golden reactants are $26.3$, $12.1$, and $30.0$, respectively.
At the reaction level, we have $79.8\%$ products with correct substructures, $63.0\%$ products with all correct substructures, and $2.4\%$ ($82.2\%-79.8\%$) products with all incorrect substructures (the percentage numbers are calculated based on the size of all test data).
It is worth noting that we will have incorrect predictions if the extracted substructures are not correct.

\begin{figure}
    \centering
    \includegraphics[scale=0.8]{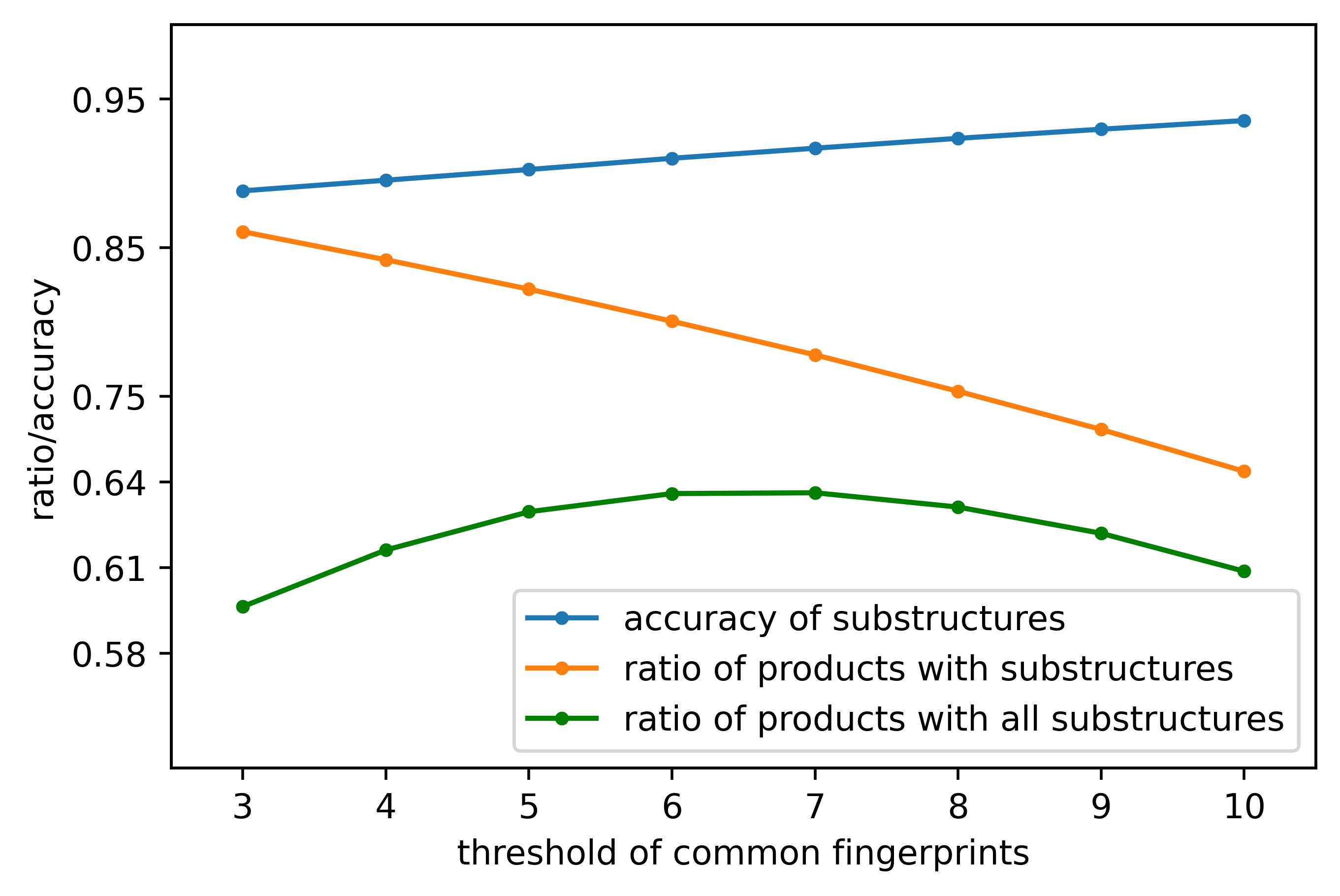}
    \caption{The accuracy and the ratio of products with substructures under different thresholds.}
    \label{fig:threshold}
\end{figure}

To improve the accuracy of extracted substructures, we plan to identify potential sites of reactivity based on the retrieved candidates, another possible way is to increase the threshold of selecting common fingerprints from retrieved candidates.
Note that we require that the common fingerprints exist in at least 5 out of the 20 retrieved candidates.
Figure~\ref{fig:threshold} shows the ratio of products with (all correct) substructures and the accuracy of substructures with the threshold ranging from $3$ to $10$.
It shows that the accuracy increases, the ratio of products with substructures decreases, and the ratio of products with all correct substructures first increases and then decreases when the threshold is set to vary from $3$ to $10$. 
In this paper, we set the threshold to $5$ mainly because it has a relatively high ratio of products with substructures, and the ratio of products with all correct substructures is also reasonably high, 
On the test data, for reactions with incorrect substructures, the average number of correct substructures is $7.3$, which means that we could obtain correct predictions even if there are incorrect substructures.

The substructures are stable and remain unchanged during reactions, thus, their reactivities should be relatively low. 
We find that the extracted substructures can be roughly classified into two categories: 1) those that are located at the end of the molecules, and are already protected by other functional groups, for instance, a hydroxy group protected by a trimethylsilyl group is a common substructure across different reaction types; 2) those located in the middle of the molecules, and are usually inert alkyl chains or aromatic rings, which contain no reactive functional groups. 
The percentages of substructures with aromaticity in the top-$10$ and $20$ most frequent substructures are  $80\%$ and $70\%$, respectively.
On average, among all the substructures, $60\%$ of the atoms have aromaticity.
These numbers show that the extracted substructures have chemical interpretability.

\begin{figure}[ht]
    \begin{center}
    \includegraphics[scale=0.15]{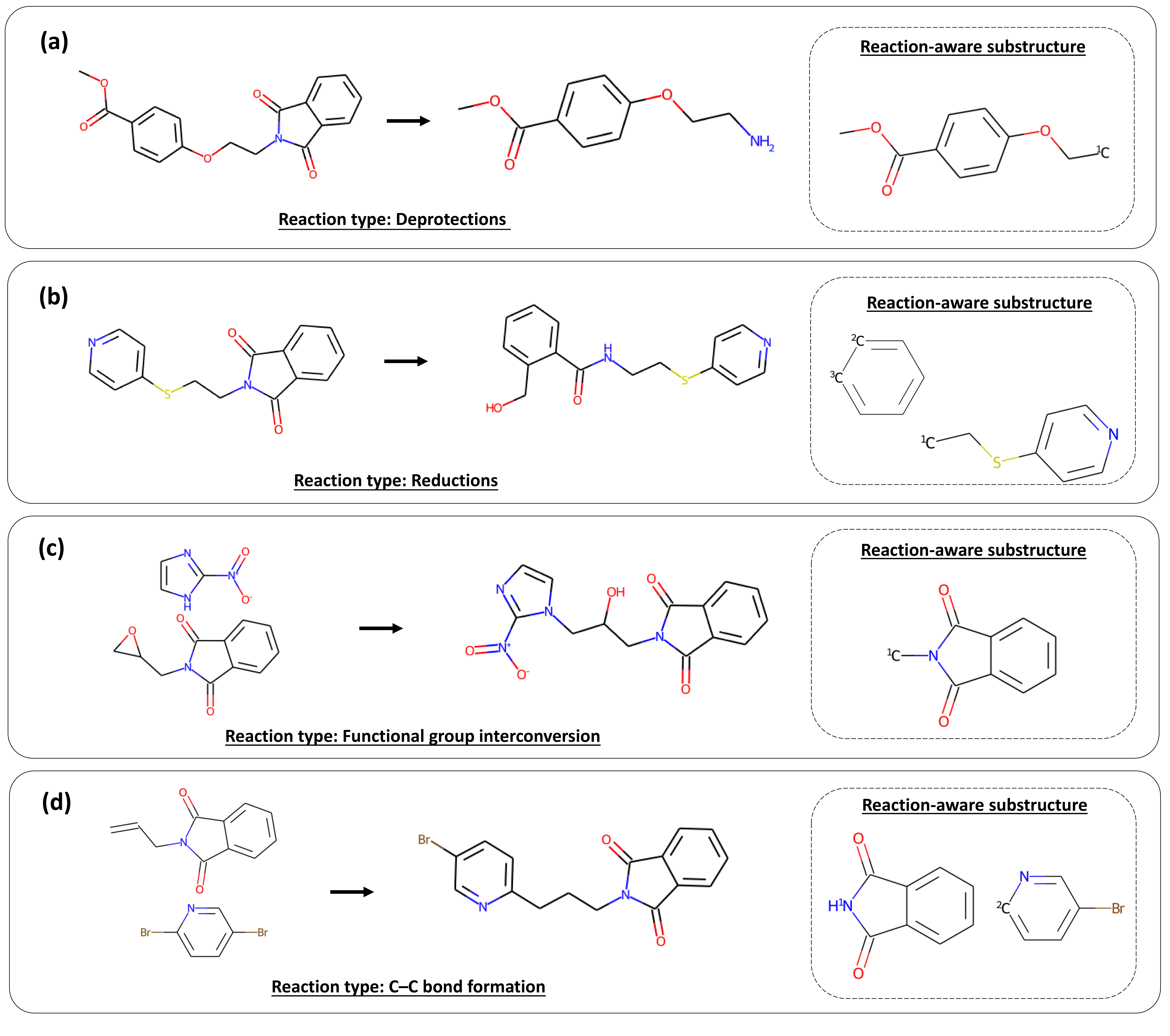}
    \caption{Reaction-aware substructures, the reactants all contain the structure of phthalimide.}
    \label{fig:rxt_aware}
    \end{center}
\end{figure}

It is important to note that the extracted substructure is reaction-aware, which captures the reaction-specific subtle chemical changes among reactants and products.
Phthalimide is a common heterocyclic substructure.
We show four exemplary reactions that all their reactants contain the structure of phthalimide in Figure~\ref{fig:rxt_aware}. 
The reaction types are also derived following\citep{schneider2015development,schneider2016big}.
The extracted substructures vary among different reaction types. 
Reaction (a) demonstrates the phthalimide as the protecting group, which has to be removed from the reactant to generate the product. 
Reaction (b) demonstrates that phthalimide is involved in reduction. 
Reaction (c) demonstrates functional group interconversion and reaction (d) shows C–C bond formation where phthalimide does not contribute atoms to the final product.
In our approach, phthalimide is not considered to be the substructure for reaction (a) and reaction (b). 
The substructures of reaction (c) and reaction (d) are also different, although they both contain phthalimide.
All these reaction-specific substructures are extracted as expected. 

Another benefit of leveraging reaction-aware substructures is that we can provide users with more insights for decision-making in retrosynthesis planning compared with existing methods.
For the case shown in Figure~\ref{fig:insight}, the product can be synthesized via multiple types of coupling reactions.
For each substructure, our approach can provide predictions and the corresponding candidates' reactions.
As is shown in Figure~\ref{fig:insight}, the reaction of the first candidate reactant is a Suzuki-Miyaura coupling reaction between the benzene and the thiophene rings, while the reaction of other candidates is a Sonogashira coupling reaction, and the end of the triple bond is the reactivity site.
It indicates that the user can refine the predictions by comparing the reactions of retrieved candidates, making our predictions more trusted compared with existing ``black-box'' models.


Note that in Figure~\ref{fig:insight}, the extracted substructure is not a fully connected graph, they come from different parts of one molecule. 
As is discussed in Section~\ref{sec:substructure}, the broken bonds might not be the sites of reactivity, for the second substructure in the bottom half of Figure~\ref{fig:insight}, all the breaking bonds are of this type.
It indicates that our approach, to some degree, is robust under different substructures, and can restore the golden targets even if the substructure is not ``perfect''.
The perfect substructure refers to a substructure that remain all the unchanged atoms during reactions.

\subsection{Results on Retrosynthesis Analysis}
\label{sec:result:retro}

\begin{table}
\begin{center}
\begin{tabular}{@{}lccccc@{}}
\toprule
Models                                     & Top-1 (\%)       & Top-10 (\%)    & \textit{Templ.}    & \textit{Map.}     \\ \midrule
RetroSim~\citep{coley2017computer}         & 32.8             & 56.1           &  \Checkmark        & \Checkmark        \\
MEGAN~\citep{sacha2021molecule}            & 33.6             & 63.9           &  \XSolidBrush        & \Checkmark      \\
GLN~\citep{dai2019retrosynthesis}          & 39.3             & 63.7           &  \Checkmark          & \Checkmark      \\ 
Graph2SMILES~\citep{tu2021permutation}     & 45.7             & 63.4           &  \XSolidBrush        & \XSolidBrush    \\
Aug. Transformer~\citep{tetko2020state}    & 44.4             & 70.4           &  \XSolidBrush        & \XSolidBrush    \\ \hline
\textbf{Ours}                              & \textbf{45.8}    & 68.2           &  \XSolidBrush        & \XSolidBrush    \\ 
\textbf{Ours} $^\ddag$                     & \textbf{47.9}    & 70.0           &                      &                 \\ \midrule \midrule
RetroPrime~\citep{wang2021retroprime}$^\dag$& 44.1            & 68.5           &  \XSolidBrush        & \Checkmark      \\
Aug.Transformer~\citep{tetko2020state}$^\dag$& 46.2           & 73.3           &  \XSolidBrush        & \XSolidBrush    \\
GTA~\citep{Seo_Song_Yang_Bae_Lee_Shin_Hwang_Yang_2021}$^\dag$ & 46.6 & 70.4    &  \XSolidBrush        & \Checkmark      \\ \hline
\textbf{Ours} $^\dag$                      & \textbf{48.0}    & 71.4           &  \XSolidBrush        & \XSolidBrush    \\ 
\textbf{Ours} $^{\ddag\dag}$               & \textbf{50.2}    & 72.9           &                      &                 \\ \bottomrule
\end{tabular}
\end{center}
\caption{The results of retrosynthesis on USPTO\_full dataset.~\textit{Templ.}: reaction templates
used;~\textit{Map.}: atom-mapping information used.
($\dag$ invalid reactions are excluded from the test set. 
$\ddag$ substructures are all correct.)}
\label{tab:retro}
\end{table}

Note that only about 80\% of training and test data has substructures, which means that our training data covers only 80\% of all the training data.
For a fair comparison, we train a vanilla Transformer model with augmented random SMILES to obtain predictions for products with no substructures.  
We report the overall results of the one-step retrosynthesis on the USPTO\_full dataset in Table~\ref{tab:retro}. 
We also add results when filtering out the predictions of all incorrect substructures as references.
For the baselines, RetroSim\citep{coley2017computer} treats retrosynthesis as template ranking based on molecular similarity, 
while MEGAN\citep{sacha2021molecule} as a sequence of molecular graph edits.
GLN\citep{dai2019retrosynthesis} employs the conditional graph logic network to learn chemical templates for retrosynthesis analysis, 
RetroPrime\citep{wang2021retroprime} decomposes the given product molecule into synthons and then generates reactants by attaching the leaving groups.
Aug.Transformer\citep{tetko2020state} incorporates data augmentation strategies with the Transformer model.
Graph2SMILES\citep{tu2021permutation} combines Transformer decoder with the permutation invariant molecular graph encoders.
GTA\citep{Seo_Song_Yang_Bae_Lee_Shin_Hwang_Yang_2021} proposes a molecular graph-aware attention mask for both self- and cross-attention in Transformer.
On the test dataset, about 4\% are invalid reactions.
We also add results with these invalid reactions excluded. 

Our method achieves a comparable or better top $1$ accuracy compared with all the baselines.
When all the substructures are all correct, the performance can be further improved, which logically indicates that putting more effort into improving the accuracy of substructures matters.
Note that our approach does not require any reaction templates built upon expert systems or template libraries, or the atom mappings from reactants to the product provided by the dataset. 
The atom-mapping information, to some degree, might reveal the information about the reactivity sites\citep{wang2021retroprime}.
The improvement in our approach can be attributed to two main factors: 
1) our approach extracts substructures for $82.2\%$ reactions on the USPTO\_full test data, a relatively high coverage, 
2) we only need to generate fragments for the isotopic labeled binding atoms in the substructures, which simplifies the problem. 
For product molecules with substructures, the average number of atoms in reactants to be predicted is reduced from $30.0$ to $17.9$.  
The design of our model has  advantages over the previous token-by-token decoding model, e.g., Aug.Transformer\citep{tetko2020state}, Graph2SMILES\citep{tu2021permutation} and GTA\citep{Seo_Song_Yang_Bae_Lee_Shin_Hwang_Yang_2021}.



\ifodd \value{is_nc}{}\else {

\section{Related Work}\label{sec:related}
The early approaches in computer-aided synthesis prediction and retrosynthesis analysis use chemical reaction rules based on subgraph pattern (reaction templates) matching, as in expert systems such as LHASA\citep{corey1972computer} and SYNTHIA\citep{szymkuc2016computer}. 
Template-based approaches use reaction templates or rule libraries, which contain reaction information about the atoms and chemical bonds near the sites of reactivity. 
Template-based methods consider all possible sites of reactivity in the molecule and enumerate possible chemical bond changes. 
These methods heavily rely on the templates, which requires considerable human effort to ensure that the template library covers most organic reactions. 

\citet{segler2017neural,coley2017prediction,baylon2019enhancing,chen2021deep} formulate reaction or retrosynthesis prediction as template classification or ranking\citep{segler2017neural,baylon2019enhancing,chen2021deep,coley2017prediction} based on molecular similarity\citep{coley2017computer} with deep neural networks. 
They select the top-ranked templates, which can then be applied to transform the input molecules into outputs. 
The templates utilized in these methods still depend on precomputed atomic mappings (how atoms in reactants map to corresponding those in products). How to obtain a complete and reliable atomic mapping relationship is also a complex problem. 

As solutions to address these limitations, several template-free approaches have been developed recently, which can be categorized into graph edit-based and translation-based approaches. 
The graph edit-based approaches cast reaction  or retrosynthesis prediction as graph transformations\citep{jin2017predicting,coley2019graph,do2019graph,bradshaw2019generative,PPR:PPR109708,sacha2021molecule}. 
Modeling or predicting electron flow in reactions\citep{bi2021non} can also be considered as a variant of graph-based methods. 
Besides, some semi-template-based methods also improve prediction performance by identifying the special sites of reactivity and then recovering graphs or sequences\citep{shi2020graph,yan2020retroxpert,NEURIPS2021_4e2a6330,wang2021retroprime}.
Translation-based approaches formalize the problems as SMILES-to-SMILES translation, typically with sequence models such as Recurrent Neural Networks\citep{nam2016linking,schwaller2018found} or the Transformer\citep{schwaller2019molecular,yang2019molecular,lin2020automatic,duan2020retrosynthesis,tetko2020state}. 
Variants of these approaches are introduced,  such as reranking and pre-training\citep{zheng2019predicting,zhu2021dual,irwin2022chemformer}. 
Some models that fuse molecule graph information with translation-based approaches also achieve promising results\citep{zhu2021dual,tu2021permutation,Seo_Song_Yang_Bae_Lee_Shin_Hwang_Yang_2021}. 

It is well accepted that substructures or functional groups are essential in chemical reactions.
\citet{wang2022chemicalreactionaware} propose new chemical-reaction-aware molecule embeddings which preserve the equivalence of reactant and product molecules in the embedding space by forcing the sum of reactant embeddings and the sum of product embeddings to be similar.
\citet{zhang2021motif} proposed motif-based graph self-supervised learning, where graph motifs refer to important subgraph patterns in molecules.
The exploration of the chemical substructure or subgraph also provides efficient solutions to build large-scale chemical libraries\citep{doi:10.1021/ci200413e} for drug discovery\citep{merlot2003chemical}. 
In our work, we explicitly introduce reaction-aware stable substructures in retrosynthesis prediction. The substructures are automatically mined with a fully data-driven approach.

\section{Conclusion}
In this paper, we introduce reaction-aware substructures to capture the subtle reactivity differences among reactants and products. 
The substructures are stable and remain unchanged during chemical reactions.
For future work, we will improve the accuracy of extracted substructures. We might also investigate how to apply the substructures to other tasks like molecule property prediction or molecular design.
}\fi

\ifodd \value{is_nc} {
\section{Data Availability} The datasets generated during and/or analysed during the current study will be available in the MCB\_SMILES repository after publication, \url{https://github.com/fangleigit/MCB\_SMILES}.
\section{Code Availability} Code and models will be available in the MCB\_SMILES repository after publication, \url{https://github.com/fangleigit/MCB\_SMILES}.
For review purpose, reviewer can view and try our live demo with the shared code and models at \url{https://github.com/fangleigit/demos}.
\section{Author contributions}
Research supervision: Jian-Guang Lou.
Lei Fang proposed to leverage reaction-aware substructures in machine learning models.
Lei Fang and Ming Zhao designed and implemented the substructure extraction algorithm.
Junren Li performed the advanced analysis on substructures.
Li Tan provided constructive suggestions and feedback, including the idea of reaction-aware substructures and manuscript writing.
\section{Competing interests}
The authors declare no competing interests.
}\else {} \fi

\bibliography{references}
\bibliographystyle{unsrtnat}
\end{document}